\documentclass[a4paper,conference]{IEEEtran}
\IEEEoverridecommandlockouts

\usepackage{cite}
\usepackage{amsmath,amssymb,amsfonts}
\usepackage{algorithm}
\usepackage{algorithmic}
\usepackage{graphicx}
\usepackage{textcomp}
\usepackage{xcolor}
\usepackage{acronym}
\usepackage{multicol}
\usepackage{subcaption}

\def\BibTeX{{\rm B\kern-.05em{\sc i\kern-.025em b}\kern-.08em
    T\kern-.1667em\lower.7ex\hbox{E}\kern-.125emX}}
    
\newacro{DoF}{degrees of freedom}
\newacro{URA}{uniform rectangular array}
\newacro{RIS}{Reconfigurable Intelligent Surface}
\newacro{CSI}{Channel State Information}
\newacro{SVD}{singular value decomposition}
\newacro{LoS}{line-of-sight}
\newacro{NLoS}{non-line-of-sight}
\newacro{IIoT}{Industrial Internet of Things}
\newacro{IRE}{intelligent reconfigurable environment}
\newacro{LIS}{Large Intelligent Surface}
\newacro{MIMO}{multiple-input multiple-output}
\newacro{SNR}{signal-to-noise ratio}
\newacro{PGM}{projected gradient method}

\DeclareMathOperator{\diag}{diag}
\DeclareMathOperator{\Tr}{Tr}
\DeclareMathOperator{\maximize}{maximize}
\DeclareMathOperator{\st}{subject~to}

\begin{document}

\title{On the Degrees of Freedom of RIS-Aided Holographic MIMO Systems\\

\thanks{J. C. Ruiz-Sicilia and M. Di Renzo are with Universit\'e Paris-Saclay, CNRS, CentraleSup\'elec, Laboratoire des Signaux et Syst\'emes, 91192 Gif-sur-Yvette, France. (juan-carlos.ruiz-sicilia@centralesupelec.fr). X. Qian was with Universit\'e Paris-Saclay, CNRS, CentraleSup\'elec, Laboratoire des Signaux et Syst\'emes, 91192 Gif-sur-Yvette, France, when this work was done. V. Sciancalepore is with NEC Laboratories Europe GmbH, Germany. M. Debbah is with the Technology Innovation Institute, Abu Dhabi, United Arab Emirates. X. Costa-P\'erez is with the i2cat Research Center, Spain, with the Catalan Institution for Research and Advanced Studies (ICREA), Spain, and with NEC Laboratories Europe GmbH, Germany. This work was supported in part by the European Commission through the H2020 MSCA 5GSmartFact project under grant agreement number 956670, the NEC Student Research Fellowship program, the H2020 ARIADNE project under grant agreement number 871464, and the H2020 RISE-6G project under grant agreement number 101017011.}
}

\author{Juan Carlos Ruiz-Sicilia, Xuewen Qian, Marco~Di~Renzo, Vincenzo Sciancalepore,\\ Merouane Debbah, and Xavier Costa-Perez}

\maketitle

\begin{abstract}
In this paper, we study surface-based communication systems based on different levels of  channel state information for system optimization. We analyze the  system performance in terms of rate and degrees of freedom (DoF). We show that the deployment of a reconfigurable intelligent surface (RIS) results in increasing the number of DoF, by extending the near-field region. Over Rician fading channels, we show that an RIS can be efficiently optimized only based on the positions of the transmitting and receiving surfaces, while providing good performance if the Rician fading factor is not too small.
\end{abstract}


\section{Introduction}
The fifth generation (5G) of wireless networks is being deployed providing improved system performance. However, the development of emerging applications, such as the \ac{IIoT}, requires ever more demanding performance in terms of rate, reliability, and number of users/devices to serve \cite{9482474}. Experimental trials and system-level simulations have shown that the performance of current wireless systems, based on optimizing only the transmitters and receivers, can be further improved by considering the environment as an additional optimization variable \cite{9038267, 9749219, 9852389}. Motivated by these considerations, a new paradigm named \ac{IRE} \cite{IRE, SmartRadioE, RenzoS19} has emerged as a new approach to overcome current design principles. In an \ac{IRE}, the propagation of the electromagnetic waves can be optimized in order to make the wireless channel between transmitters and receivers more reliable, while reducing the implementation complexity of transmitters and receivers. The control of the environment is carried out by using transmitting, receiving, and reflecting surfaces based on programmable metamaterials. These surfaces can be either active or nearly-passive. Active surfaces that operate as transmitters and receivers are referred to as holographic surfaces (HoloS) \cite{9136592}. Nearly-passive surfaces that operate as, e.g., reflecting or refracting surfaces, are referred to as reconfigurable intelligent surfaces (RIS) \cite{DBLP:journals/ejwcn/RenzoDHZAYSAHGR19}, \cite{9326394}.

HoloS can be utilized as flexible antennas that are electrically large. Their large size compared to the wavelength increases the Fraunhofer distance, making the plane wave far-field assumption no longer valid in many scenarios, especially for operation at very high frequency bands \cite{DBLP:journals/corr/abs-2210-08616}. In this case, the wavefront of the electromagnetic waves is not planar anymore, which opens new communication opportunities, e.g., the possibility of spatial multiplexing even in \ac{LoS} \ac{MIMO} channels \cite{Tse}, \cite{Dardari_2020}. RIS can be utilized, on the other hand, to establish strong transmission links by appropriately shaping the signals reflected or refracted by existing material objects, in order to, e.g., solve coverage hole problems \cite{DBLP:journals/ejwcn/RenzoDHZAYSAHGR19}, \cite{9326394}. The combination of HoloS and RIS enables the transmission through high-rank channels ensuring high \ac{SNR} links, eventually leading to high capacity gains. 

In this paper, motivated by these considerations, we analyze different strategies to optimize an RIS-aided HoloS communication system as a function of the level of channel state information (CSI) for optimizing the RIS. In particular, we focus our attention on the achievable number of \ac{DoF} under different scenarios \cite{DBLP:journals/corr/abs-2210-08616}. To this end, the paper is organized as follows. Section \ref{sec:systemModel} introduces the surface-based communication system model. Section \ref{sec:systemArchitecture} summarizes the considered RIS configurations as a function of the available CSI. Section \ref{sec:results} presents some numerical results to compare the rate and the \ac{DoF} of the considered case studies. Finally, Section \ref{sec:conclusions} concludes the paper.

\textit{Notation}: Bold lower and upper case letters represent vectors and matrices, respectively. $\mathbb{C}^{a \times b}$ denotes the space of complex matrices of dimensions $a \times b$. $(\cdot)^*$ and $(\cdot)^H$ represent the complex conjugate and the Hermitian transpose. $\diag(\mathbf{x})$ denotes the square diagonal matrix which has the elements of $\mathbf{x}$ on the main diagonal. $\Tr(\mathbf{X})$ is the trace of matrix $\mathbf{X}$, and $\mathbb{E}\{\cdot\}$ stands for the expectation operator. The notation $\mathbf{A}\succeq(\succ) \mathbf{B}$ means that $\mathbf{A}-\mathbf{B}$ is positive semidefinite (definite). $\nabla_{\mathbf{X}} f (\cdot)$ is the gradient of $f$ with respect to $\mathbf{X}^* \in \mathbb{C}^{a \times b}$, which also lies in $\mathbb{C}^{a \times b}$. $\mathbf{A}(i,k)$ denotes the $k$-th element of the $i$-th row of matrix $\mathbf{A}$. $j$ is the imaginary unit. $\mathcal{CN}(\Bar{x},\sigma^2)$ denotes the complex Gaussian distribution with mean $\Bar{x}$ and variance $\sigma^2$.

\section{System model}
\label{sec:systemModel}
\begin{figure}[!t]
    \centering
    \includegraphics[width=\columnwidth]{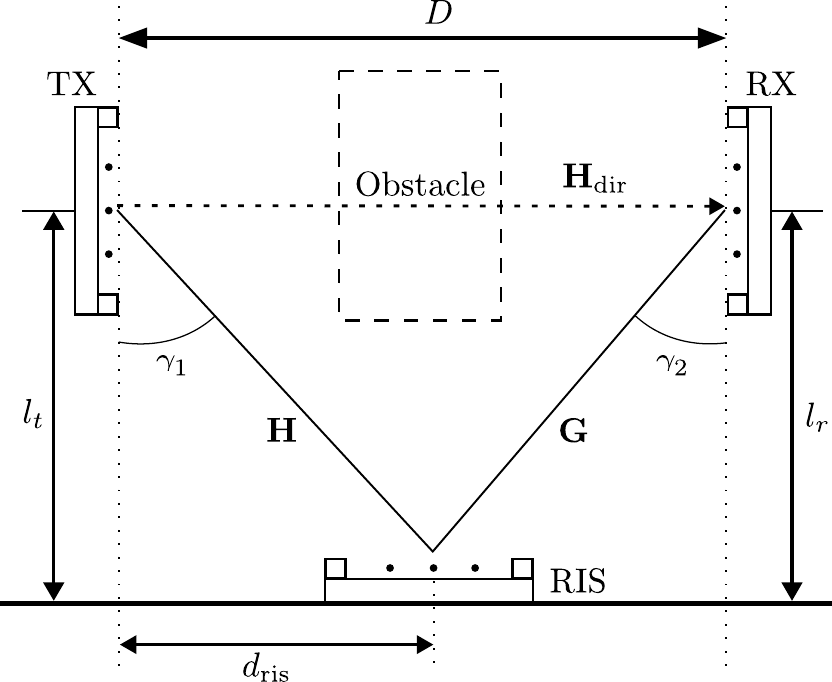}
    \caption{Top view of the considered communication system model \cite{Perovic}.}
    \label{fig:system_model} \vspace{-0.11cm}
\end{figure}

We consider an RIS-aided HoloS communication system as illustrated in Fig. \ref{fig:system_model}. The transmitter and receiver are modeled as \acp{URA} with $L= L_y L_z$ and $M = M_y M_z$ antenna elements, respectively, where $L_y$ ($M_y$) and $L_z$ ($M_z$) are the numbers of antenna elements on the $y$-axis and $z$-axis. $G_t$ and $G_r$ denote the gains of the antenna elements at the transmitter and receiver, respectively. The transmitter and receiver are placed on vertical walls parallel to each other. $D$ denotes the distance between these walls. 

The RIS consists of $N_x$ unit cells on the $x$-axis and $N_y$ unit cells on the $y$-axis (i.e., $N = N_x N_y$ is the total number of unit cells). The RIS is installed on a vertical wall that is perpendicular to the HoloS transmitter and HoloS receiver, and the distance between the center of the RIS and the plane containing the HoloS is $d_{\mathrm{ris}}$. The distance between the center of the transmitting HoloS and the plane containing the RIS is $l_{t}$, and the distance between the midpoint of the receiving HoloS and the plane containing the RIS is $l_{r}$. 

For simplicity, we assume that the three surfaces are at the same height. In the three surfaces, the centers between adjacent elements are separated by $d=\lambda/2$ to avoid the mutual coupling among them, where $\lambda$ is the wavelength. Thus, the area of each element is $S_c = \lambda^2/4$. The position of the $l$-th transmitting element, the $m$-th receiving element, and the $n$-th unit cell of the RIS are denoted by $\mathbf{r}^l_t = (x^l_t,y^l_t,z^l_t)$, $\mathbf{r}^m_r = (x^m_r,y^m_r,z^m_r)$ and $\mathbf{r}^n_\mathrm{ris} = (x^n_\mathrm{ris},y^n_\mathrm{ris},z^n_\mathrm{ris} )$, respectively. The positions can be formulated as follows:
\begin{equation}
    \mathbf{r}^l_t = \left(-d_{\mathrm{ris}},d l_y - \frac{d}{2}(L_y+1),l_t + d l_z - \frac{d}{2}(L_z+1) \right)
    \end{equation}
\begin{multline}
    \mathbf{r}^m_r = \left(\vphantom{\frac12} D-d_{\mathrm{ris}},d m_y - \frac{d}{2}(M_y+1),
    \right.\\ \left.
    l_r + d m_z - \frac{d}{2}(M_z+1) \right)
\end{multline}
\begin{equation}
    \mathbf{r}^n_\mathrm{ris} = \left( n_x d-\frac{d}{2}(N_{x}+1), n_yd-\frac{d}{2}(N_{y}+1), 0 \right)
\end{equation}
where $l = (l_z-1)L_z+l_y$, $m = (m_z-1)M_z+m_y$ and $n= (n_x-1)N_x+n_y$, and $l_z$ ($m_z$) and $l_y$ ($m_y$) denote the indices of the \acp{URA} along the $z$-axis and $y$-axis, respectively. Similarly, $n_x$ and $n_y$ denote the indices of the unit cells of the RIS along the $x$-axis and $y$-axis.

As for the RIS, we consider the canonical reflectarray-based model \cite{9864116}, according to which the reflection coefficients of the $N$ unit cells are denoted by the vector $\boldsymbol{\theta} = \diag([e^{j\theta_1},e^{j\theta_2},...,e^{j\theta_N}])$, where $\theta_n \in [-\pi, \pi]$ is the phase shift applied by the $n$-th unit cell. The direct channel between the transmitting and receiving HoloS is  denoted by $\mathbf{H}_\mathrm{dir} \in \mathbb{C}^{M \times L}$. The channel from the transmitting HoloS to the RIS and the channel from the RIS to the receiving HoloS are denoted by $\mathbf{H} \in \mathbb{C}^{N \times L}$ and $\mathbf{G} \in \mathbb{C}^{M \times N}$, respectively. Then, the signal $\mathbf{y}$ at the receiver is given by:
\begin{equation}
    \mathbf{y} = (\mathbf{H}_\mathrm{dir} + \mathbf{G}\boldsymbol{\theta}\mathbf{H})\mathbf{x}+ \boldsymbol{\omega}
\end{equation}
where $\boldsymbol{\omega}$ denotes the white additive Gaussian noise with zero means and variance $ \sigma^2 $, i.e., $\mathbb{E}\{\boldsymbol{\omega}\boldsymbol{\omega}^H\} = \sigma^2\mathbf{I}_M$ and $ \mathbf{x} $ denotes the transmitted signal. The symbols in $\mathbf{x}$ are distributed according to a circularly symmetric complex Gaussian distribution with $\mathbb{E}\{\mathbf{x}\mathbf{x}^H\}=\mathbf{Q}$, where $\mathbf{Q}$ is the  covariance matrix. Assuming that the maximum transmit power is $P_T$, we obtain $\Tr(\mathbf{Q}) \leq P_T$. 

Given this system model, the achievable rate can be written as:
\begin{multline}
\label{eq:capacity}
    R(\boldsymbol{\theta}, \mathbf{Q}) =\\ \log_2\left|\mathbf{I} + \frac{ (\mathbf{H}_\mathrm{dir} + \mathbf{G}\boldsymbol{\theta}\mathbf{H}) \mathbf{Q} (\mathbf{H}_\mathrm{dir} + \mathbf{G}\boldsymbol{\theta}\mathbf{H})^H }{\sigma^2 } \right|
\end{multline}

\subsection{Channel Model}
The links between each pair of surfaces are modeled as Rician channels with Rician factor $K$. Hence, the channels can be formulated as follows:
\begin{align}
    \mathbf{G} = \sqrt{\frac{K}{K+1}}\mathbf{G}^{\mathrm{LoS}} + \sqrt{\frac{1}{K+1}}\mathbf{G}^{\mathrm{NLoS}} \\
    \mathbf{H} = \sqrt{\frac{K}{K+1}}\mathbf{H}^{\mathrm{LoS}} + \sqrt{\frac{1}{K+1}}\mathbf{H}^{\mathrm{NLoS}}
\end{align}
where $\mathbf{G}^{\mathrm{LoS}}$ ($ \mathbf{H}^{\mathrm{LoS}} $) and $ \mathbf{G}^{\mathrm{NLoS}} $ ($ \mathbf{H}^{\mathrm{NLoS}} $) denote the \ac{LoS} and the \ac{NLoS} components of each channel, respectively. 

The \ac{LoS} component of the channel from the $l$-th element of the transmitting HoloS to the $n$-th unit cell of the RIS can be formulated as \cite[Eq. (5)]{Pathloss}:
\begin{equation}
\label{eq:hNear}
    \mathbf{H}^{\mathrm{LoS}}(n,l) = \sqrt{ \frac{G_t S_c }{4\pi}\frac{\cos \gamma_1}{(d_1(n,l))^{2}} } e^{jk_0 d_1(n,l)}
\end{equation}
where $G_c = S_c 4 \pi/\lambda^2$, $k_0 = 2\pi/\lambda$, $\cos \gamma_1 = z_t^l/d_1(n,l)$ and
\begin{equation}
\label{eq:d1Exact}
    d_1(n,l) = \sqrt{ (x^l_t-x^n_\mathrm{ris})^2+(y^l_t-y^n_\mathrm{ris})^2+(z^l_t-z^n_\mathrm{ris})^2 }
\end{equation}

Similarly, the \ac{LoS} component of the channel from the $n$-th unit cell of the RIS to the $m$-th element of the receiving HoloS can be expressed as follows \cite[Eq. (6)]{Pathloss}:
\begin{equation}
    \mathbf{G}^{\mathrm{LoS}}(m,n) = \sqrt{ \frac{G_r S_c }{4\pi}\frac{\cos \gamma_2}{(d_2(m,n))^{2}} } e^{j k_0 d_2(m,n)}
\end{equation}
\noindent
where $\cos \gamma_2 = z_r^m/d_2(m,n)$ and
\begin{multline}
    d_2(m,n) =\\ \sqrt{ (x^r_l-x^n_\mathrm{ris})^2+(y^r_l-y^n_\mathrm{ris})^2+(z^r_l-z^n_\mathrm{ris})^2 }
\end{multline}

The \ac{NLoS} components of the channels can be formulated as follows:
\begin{align}
    \mathbf{G}^{\mathrm{NLoS}}(m,n) &= | \mathbf{G}^{\mathrm{LoS}}(m,n)| x_{m,n} \\
    \mathbf{H}^{\mathrm{NLoS}}(n,l) &= | \mathbf{H}^{\mathrm{LoS}}(n,l)| x_{n,l}
\end{align}
where $x_{m,n}$ and $x_{n,l}$ are independent and identically distributed standard Gaussian random variables, i.e., $ x_{m,n}, x_{n,l} \sim \mathcal{CN}(0,1)$. 

As far as the direct channel is concerned, we have:
\begin{equation}
    \mathbf{H}_\mathrm{dir} = \sqrt{\frac{K}{K+1}}\mathbf{H}_\mathrm{dir}^{\mathrm{LoS}} + \sqrt{\frac{1}{K+1}}\mathbf{H}_\mathrm{dir}^{\mathrm{NLoS}}
\end{equation}
where $\mathbf{H}_\mathrm{dir}^{\mathrm{LoS}}$ and $ \mathbf{H}_\mathrm{dir}^{\mathrm{NLoS}}$ denote the \ac{LoS} and \ac{NLoS} components, respectively. 

The \ac{LoS} component can be written as follows:
\begin{equation}
    \mathbf{H}_\mathrm{dir}(m,l) = \sqrt{ \frac{G_t G_r \lambda^2}{(4\pi)^2 (d_\mathrm{dir}(n,m))^{\alpha_{\mathrm{dir}}}}} e^{jk_0 d_\mathrm{dir}(n,m)}
\end{equation}
where $\alpha_{\mathrm{dir}}$ is the path loss exponent and
\begin{equation}
    d_\mathrm{dir}(n,m) = \sqrt{ (x^l_t-x^m_r)^2+(y^l_t-y^m_r)^2+(z^l_t-z^m_r)^2 }
\end{equation}

Similarly, the \ac{NLoS} component $\mathbf{H}_\mathrm{dir}^{\mathrm{NLoS}}$ can be formulated as:
\begin{equation}
    \mathbf{H}_\mathrm{dir}^{\mathrm{NLoS}} (m,l) = |\mathbf{H}_\mathrm{dir}^{\mathrm{LoS}}(m,l)| x_{m,l}
\end{equation}
where $ x_{m,l} \sim \mathcal{CN}(0,1) $.

\section{Problem Formulation}
\label{sec:systemArchitecture}
In this section, we consider several case studies for optimizing the transmitting and receiving HoloS, and the RIS as a function of the CSI that is needed for optimizing the RIS. On the other hand, we assume that the transmitting and receiving HoloS have perfect CSI. It is known, in fact, that the bottleneck when optimizing RIS-aided systems is given by the overhead for configuring the RIS because of its nearly-passive implementation \cite{9847080}.

To make the problem formulation parametric as a function of the CSI assumed for optimizing the RIS, the objective function in \eqref{eq:capacity} is rewritten as a function of three generic channel matrices $\mathbf{F}_1$, $\mathbf{F}_2$ and $\mathbf{F}_3$, as follows:
\begin{multline}
\label{eq:objFunction}
    f(\boldsymbol{\theta}, \mathbf{Q}; \mathbf{F}_1, \mathbf{F}_2, \mathbf{F}_3) =\\ \log_2\left|\mathbf{I} + \frac{ (\mathbf{F}_1 + \mathbf{F}_3 \boldsymbol{\theta} \mathbf{F}_2) \mathbf{Q} (\mathbf{F}_1 + \mathbf{F}_3 \boldsymbol{\theta} \mathbf{F}_2))^H }{\sigma^2 } \right|
\end{multline}

The three matrices $\mathbf{F}_1$, $\mathbf{F}_2$ and $\mathbf{F}_3$ are specialized for each considered case study in the following subsections. With the notation in \eqref{eq:objFunction}, we mean that the phase shifts of the RIS are optimized assuming that $\mathbf{F}_1$, $\mathbf{F}_2$ and $\mathbf{F}_3$ are known.

Based on \eqref{eq:objFunction}, the optimization problem of interest can be stated as follows:
\begin{subequations}\label{eq:capacityprob}
\begin{align}
\underset{\boldsymbol{\theta},\mathbf{Q}}{\maximize}
& \ f(\boldsymbol{\theta}, \mathbf{Q}; \mathbf{F}_1,  \mathbf{F}_2, \mathbf{F}_3) \label{eq:opt_problem}\\
\st & \ \Tr(\mathbf{Q})\le
P_{t};\mathbf{Q}\succeq\mathbf{0};\\
 & \ \bigl|\boldsymbol{\theta}(n,n)\bigr|=1,n=1,2,\ldots,N
\end{align}
\end{subequations}

In \cite{Perovic}, the authors introduce an algorithm to solve the problem in (19) based on the \ac{PGM}. The approach proposed in \cite{Perovic} employs one step size to optimize $\mathbf{Q}$ and $\boldsymbol{\theta}$ in an iterative manner. The proposed algorithm needs a careful choice of the step size. To this end, the authors introduce a scaling factor that needs to be finely tuned depending on the considered scenario. A more efficient solution is proposed in \cite{cutoff}, where the authors generalize the approach in \cite{Perovic} by using two different step sizes for optimizing $\boldsymbol{\theta}$ and $\mathbf{Q}$. This approach avoids the need of finely tuning an ad hoc scaling factor for each considered scenario.

Inspired by \cite{cutoff}, we propose the following iterative algorithm to solve the problem in (19):
\begin{subequations}
\begin{align}
\boldsymbol{\theta}_{n+1}  &=P_{\Theta}(\boldsymbol{\theta}_{n}+ \mu_1 \nabla_{\boldsymbol{\theta}} f(\boldsymbol{\theta}_{n}, \mathbf{Q}_{n}; \mathbf{F}_1, \mathbf{F}_2, \mathbf{F}_3))\\
\mathbf{Q}_{n+1} &= P_{\mathcal{Q}}(\mathbf{Q}_{n} + \mu_2 \nabla_{\mathbf{Q}} f(\boldsymbol{\theta}_{n+1}, \mathbf{Q}_{n}; \mathbf{F}_1, \mathbf{F}_2, \mathbf{F}_3))
\end{align}
\end{subequations}
where $P_{\Theta}(\cdot)$ and $P_{\mathcal{Q}}(\cdot)$ denote the projection onto the feasible sets of $\boldsymbol{\theta}$ and $\mathbf{Q}$, respectively. The gradients and the projections can be found in \cite[Eqs. (17a), (17b), (18), (21)]{Perovic}.

In (20), $\mu_1$ and $\mu_2$ denote the step sizes, which need to be carefully chosen in order to ensure convergence. For this purpose, we implement the Armijo-Goldstein backtracking line search to adjust the largest possible step size in each iteration of the algorithm \cite{cutoff}. To this end, we define $L_1, L_2 >0$ as the maximum initial step sizes and $\rho_1, \rho_2 \in (0,1)$ as the search control parameters. The backtracking line search method consists of replacing $\mu_1$ and $\mu_2$ with $L_1 \rho_1^{\alpha_n}$ and $L_2 \rho_2^{\beta_n}$, respectively, where $\alpha_n$ and $\beta_n$ are the minimum non-negative integer values fulfilling the conditions:
\begin{multline}
    f(\boldsymbol{\theta}_{n+1}, \mathbf{Q}_{n}; \mathbf{F}_1, \mathbf{F}_2, \mathbf{F}_3) \leq \\
    f(\boldsymbol{\theta}_{n}, \mathbf{Q}_{n}; \mathbf{F}_1, \mathbf{F}_2, \mathbf{F}_3)- \delta_1 \left \lVert \boldsymbol{\theta}_{n+1} - \boldsymbol{\theta}_{n}\right \lVert^2
\end{multline}
\begin{multline}
    f(\boldsymbol{\theta}_{n+1}, \mathbf{Q}_{n+1}; \mathbf{F}_1, \mathbf{F}_2, \mathbf{F}_3) \leq\\
    f(\boldsymbol{\theta}_{n+1}, \mathbf{Q}_{n}; \mathbf{F}_1, \mathbf{F}_2, \mathbf{F}_3) - \delta_2 \left \lVert \mathbf{Q}_{n+1} - \mathbf{Q}_{n}\right \lVert^2
\end{multline}
with being $\delta_1, \delta_2 > 0$ small constants.

\subsection{Scheme 1: Optimization Based on Perfect CSI}
This is the benchmark scheme (optimal configuration), where it is assumed that perfect CSI is available for optimizing the RIS, i.e., $\mathbf{H}_\mathrm{dir}$, $\mathbf{H}$ and $\mathbf{G}$ are assumed to be known. The optimal values of $\boldsymbol{\theta}^{\rm{opt}}$ and $\mathbf{Q}^{\rm{opt}}$ are obtained solving \eqref{eq:capacityprob} with the proposed algorithm and by setting $\mathbf{F}_1 = \mathbf{H}_\mathrm{dir}$, $\mathbf{F}_2 = \mathbf{H}$, $\mathbf{F}_3 = \mathbf{G}$.

\subsection{Scheme 2: Optimization Based on LoS CSI}
In this case, we assume that the RIS is optimized only based on the prior knowledge of the \ac{LoS} components of the channels, i.e., $\mathbf{H}_\mathrm{dir}^{\mathrm{LoS}}$, $\mathbf{H}^{\mathrm{LoS}}$ and $\mathbf{G}^{\mathrm{LoS}}$. Specifically, the optimization is carried out in two steps. 

Firstly, the RIS vector of the phase shifts of the RIS is optimized by solving \eqref{eq:capacityprob} with the proposed algorithm and setting $\mathbf{F}_1 = \mathbf{H}_\mathrm{dir}^{\mathrm{LoS}}$, $\mathbf{F}_2 = \mathbf{H}^{\mathrm{LoS}}$, $\mathbf{F}_3 = \mathbf{G}^{\mathrm{LoS}}$. The obtained solution is denoted as $\boldsymbol{\theta}^\mathrm{LoS}$. It is worth mentioning that in this step the covariance matrix $\mathbf{Q}$ is optimized as well. However, the obtained matrix is disregarded. 

Secondly, once the RIS is optimized, the end-to-end channel is $\mathbf{\tilde{H}} = \mathbf{H}_\mathrm{dir} + \mathbf{G}\boldsymbol{\theta}^\mathrm{LoS}\mathbf{H}$. Based on this channel, the covariance matrix $\mathbf{Q}$ can be obtained by applying the water-filling solution \cite{Tse}, under the assumption that all the channels are perfectly known at the transmitting and receiving HoloS. By applying the \ac{SVD} to the end-to-end channel $\mathbf{\tilde{H}}$, we obtain $\mathbf{\tilde{H}} = \mathbf{U}\mathbf{S}\mathbf{V}^H$ with $S = \diag(\lambda_1,\lambda_2,\ldots,\lambda_{N_s})$, $\mathbf{V} \in \mathbb{C}^{L \times N_s}$, and $N_s \leq \min(M,L)$. Then, the optimal $\mathbf{Q}^\mathrm{LoS}$ is given by:
\begin{equation}
    \mathbf{Q}^\mathrm{LoS} = \mathbf{V} \diag(P_{T,1},P_{T,2},\ldots, P_{T,N_s})\mathbf{V}^H
\end{equation}
where $P_{T,k}$ is the power allocated to the $k$-th communication mode, which is given by $P_{T,k} = \max(\mu - \frac{\sigma^2}{\lambda_k^2},0)$ with $\mu$ chosen to fulfill the condition  $\sum_{k=1}^{k=N_s} P_{T,k} = P_T$.

\subsection{Scheme 3: Optimization Based on Location Information}
This case study is motivated by the focusing function introduced in \cite{Miller} and, later, studied in \cite{Bartoli} and \cite{9973178} for RIS-aided LoS channels. Similar to Scheme 3, we assume that the RIS is optimized only based on the LoS components of the channel. However, we further relax the prior knowledge for optimizing the RIS and assume that $\boldsymbol{\theta}$ is optimized only based on the knowledge for the center positions of the transmitting and receiving HoloS.

Specifically, inspired by the concept of focusing function defined in \cite{Miller}, the channel phase of the direct channel between the center points of the transmitting and receiving HoloS can be written as follows:
\begin{equation}
    \Phi_\mathrm{dir} = k_0 d_\mathrm{dir} = k_0 \sqrt{D^2 + (l_t-l_r)^2}
\end{equation}

Likewise, the accumulated channel phase of the link between the center point of the transmitting HoloS and the $n$-th unit cell of the RIS and from the $n$-th unit cell of the RIS to the center point of the receiving HoloS can be expressed as:
\begin{equation}
     \Phi^\mathrm{focus}_\mathrm{indir}(n) = k_0 d_{1}^\mathrm{focus}(n) + \boldsymbol{\theta}(n,n) + k_0 d_{2}^\mathrm{focus}(n)
\end{equation}
where
\begin{align}
    d_{1}^\mathrm{focus}(n) &= \sqrt{ (-d_{\mathrm{ris}}-x^{\mathrm{ris}}_n)^2+(y^{\mathrm{ris}}_n)^2+ l_t^2 } \label{eq:d1Focus}\\
    d_{2}^\mathrm{focus}(n) &= \sqrt{ (D-d_{\mathrm{ris}}-x^{\mathrm{ris}}_n)^2+(y^{\mathrm{ris}}_n)^2+ l_r^2 } \label{eq:d2Focus}
\end{align}

In this case study, we assume that the phase shifts of the RIS are optimized to compensate for the phase shifts $\Phi^\mathrm{focus}_\mathrm{indir}(n)$, so that all the reflected signals reach the receiving HoloS with the same phase as the direct channel, i.e., $\Phi_\mathrm{dir} = \Phi^\mathrm{focus}_\mathrm{indir}(n)$ for $n\in [1,N]$. Hence, the $n$-th element of the phase shift vector of the RIS can be formulated as follows:
\begin{equation}
    \boldsymbol{\theta}^\mathrm{focus}(n,n) = k_0( d_\mathrm{dir} - d_{1}^\mathrm{focus}(n) - d_{2}^\mathrm{focus}(n))
\end{equation}

Once the RIS is optimized, the end-to-end channel is given and known. Thus, $\mathbf{Q}$ is optimized as for Scheme 2.

\subsection{Scheme 4: Far-Field Optimization (Anomalous Reflection)}
In this case study, we optimize $\boldsymbol{\theta}$ and $\mathbf{Q}$ as for Scheme 3 under the assumption that the transmitting and receiving HoloS are in the far-field of each other and of the RIS.

Therefore, the distances in \eqref{eq:d1Focus} and \eqref{eq:d2Focus} can be simplified. Let us consider \eqref{eq:d1Focus}. We can rewrite it as follows: 
\begin{equation}
\begin{split}
    d_{1}^\mathrm{focus}(n) &= \sqrt{ d_{\mathrm{ris}}^2 + 2 d_{\mathrm{ris}} x^n_\mathrm{ris} + (x^n_\mathrm{ris})^2 + (y^n_\mathrm{ris})^2 + l_t^2}\\
    &= d_1 \sqrt{1 + \frac{2 d_{\mathrm{ris}} x^n_\mathrm{ris} + (x^n_\mathrm{ris})^2 + (y^n_\mathrm{ris})^2}{d_1^2} }
\end{split}
\label{eq:d1Focus_bis}
\end{equation}
where $d_1 = \sqrt{d_{\mathrm{ris}}^2+l_t^2}$. 

Under the assumption of far-field propagation, we have:
\begin{equation}
    d_1 \gg x^n_\mathrm{ris}, y^n_\mathrm{ris}
\end{equation}
which results in:
\begin{equation}
\label{eq:d1_far}
    d_{1}^\mathrm{far}(n) \approx d_1 + \frac{ d_{\mathrm{ris}} x^n_\mathrm{ris}}{d_1}
\end{equation}

By applying the same approximation to \eqref{eq:d2Focus}, we obtain:
\begin{equation}
\label{eq:d2_far}
    d_{2}^\mathrm{far}(n) \approx d_2 - \frac{ (D-d_{\mathrm{ris}}) x^n_\mathrm{ris}}{d_2}
\end{equation}
where $d_2 = \sqrt{(D-d_{\mathrm{ris}})^2+l_r^2}$.

By using the same line of thought as for Scheme 3, we have:
\begin{equation}
\label{eq:RIS_far}
\boldsymbol{\theta}^\mathrm{far}(n,n) = k_0( d_\mathrm{dir} - d_{1}^\mathrm{far}(n) - d_{2}^\mathrm{far}(n)).
\end{equation}

Once the RIS is optimized, the end-to-end channel is given and known. Thus, $\mathbf{Q}$ is optimized as for Scheme 2 and Scheme 3.

\section{Numerical Results}
\label{sec:results}
In this section, we illustrate some numerical results to analyze the performance of the considered case studies. We assume a setup in which the transmitting and receiving HoloS are equipped with $8 \times 8$ elements. The simulation parameters are as follows: $l_t = 2$ m, $l_r = 2$ m, $G_t = 3$ dBi, $G_r = 3$ dBi, $f = 3.5$ GHz, $P_T = -10$ dBm, and $\alpha_{\mathrm{dir}}=3$. We consider a noise spectral density $N_0 = -170$ dBm/Hz and a transmission bandwidth $BW=20$ MHz resulting in a noise variance $\sigma^2 = N_0 BW = -97$ dBm.

\begin{figure}[t]
\centering
 \includegraphics[width=0.90\columnwidth]{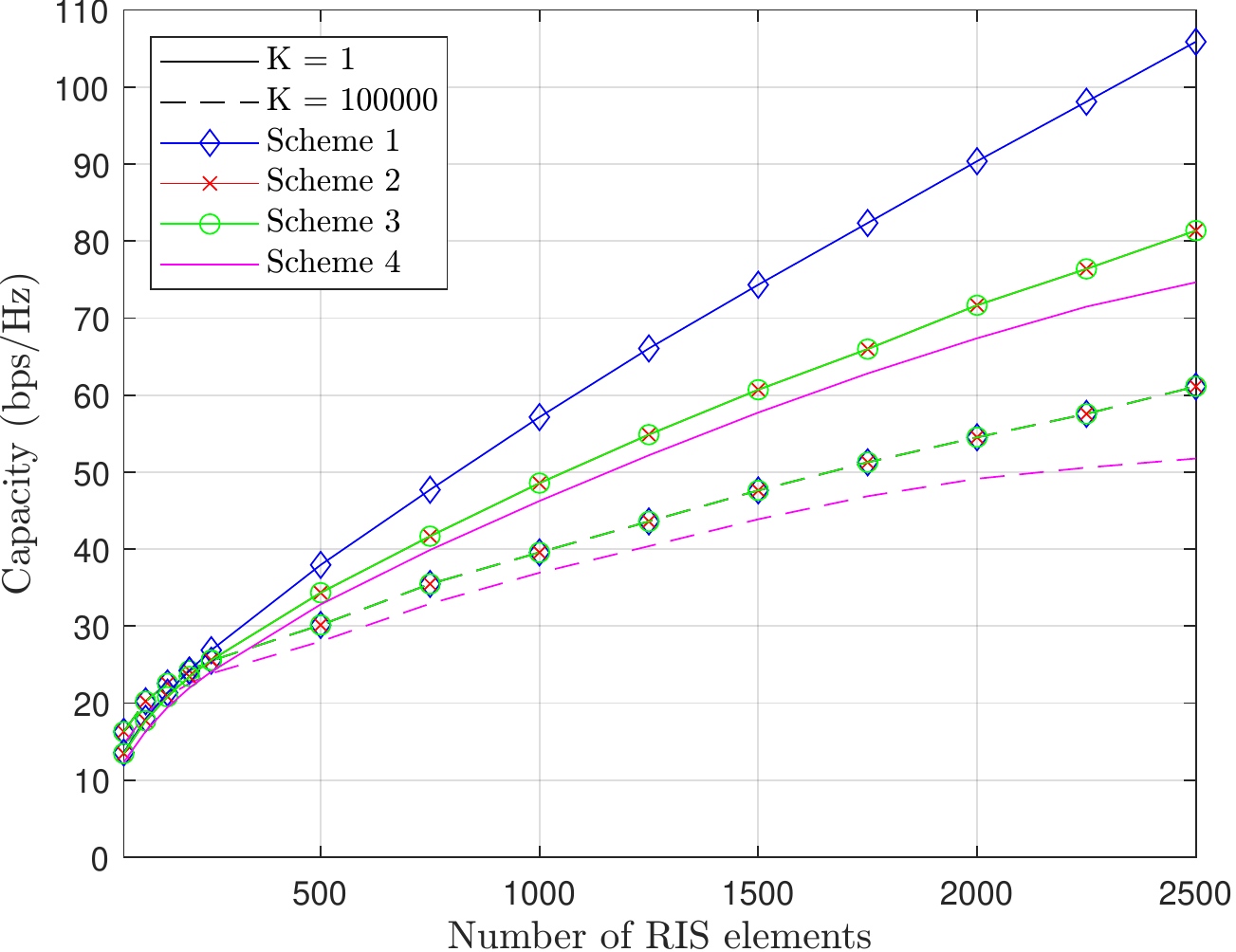}
\caption{Achievable rate as function of the number of RIS elements $N$.}
\label{fig:capacity}
\end{figure}

In Fig. \ref{fig:capacity}, we show the achievable rate as a function of the number of unit cells of the RIS, by assuming that the direct channel is blocked by an obstacle, and $d_\mathrm{ris} = D/2$ with $D = 15$ m. As expected, the rate increases for smaller values of $K$, since the channel is richer in terms of multipaths and the number of DoF increases. However, it is important to note that a high rate is obtained even in LoS conditions ($K=100,000$), thanks to the spherical wavefront under the considered setup. Specifically, we note that the rate obtained for low values of RIS elements is in agreement with the rate of a single-input single-output channel with a received signal-to-noise-ratio that corresponds to the considered setup when the transmission distances are those between the center-points of the transmitter and the RIS, and the center-points of the RIS and the receiver. Notably, in addition, we see that Scheme 3 and Scheme 1 offer almost the same performance for large values of $K$ (LoS conditions). 

\begin{figure}[t]
\centering
 \includegraphics[width=0.90\columnwidth]{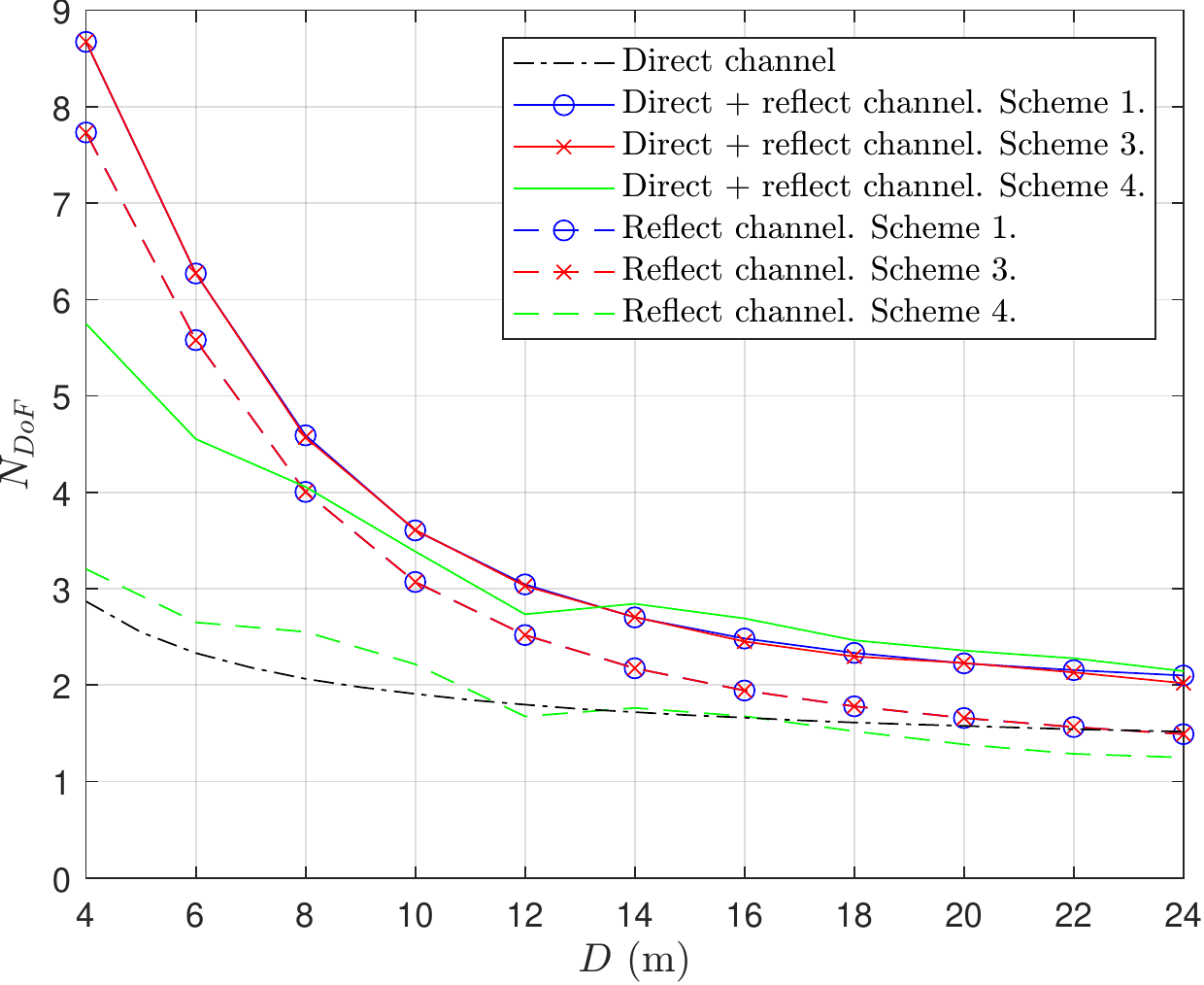}
\caption{\ac{DoF} as a function of the transmission distance.}
\label{fig:DoF_varyD} \vspace{-0.25cm}
\end{figure}

In Fig. \ref{fig:DoF_varyD}, we illustrate the number of \ac{DoF} when the size of the RIS is $50\times 50$ unit cells. To evaluate the number of strongly coupled transmission modes, we plot the effective rank \cite{effRank}. Also, we assume LoS conditions ($K=100,000$). The obtained results clearly show that the deployment of an RIS can extend the near-field region between the transmitting and receiving HoloS, to an extent that the end-to-end channel has a number of DoF that exceeds the number of DoF of the direct link, especially if the distance $D$ is small enough compared with the sizes of the transmitting and receiving HoloS and the RIS. Once again, Scheme 3 provides almost the same number of DoF as Scheme 1. As expected, the number of DoF tends to one when the transmission distance is large.

For illustrative purposes, it is interesting to visualize the shape of the optimal spatial functions (communication modes) at the transmitting HoloS when they are observed at the RIS. In mathematical terms, we plot the following:
\begin{equation}
    \mathbf{w}_i = \mathbf{H}\mathbf{v}_i \sqrt{P_{Ti}}
\end{equation}
where $\mathbf{v}_i$ corresponds to the $i$-th column of the matrix $\mathbf{V}$. 

\begin{figure*}
\centering
\begin{subfigure}{\columnwidth}
\includegraphics[width=0.90\columnwidth]{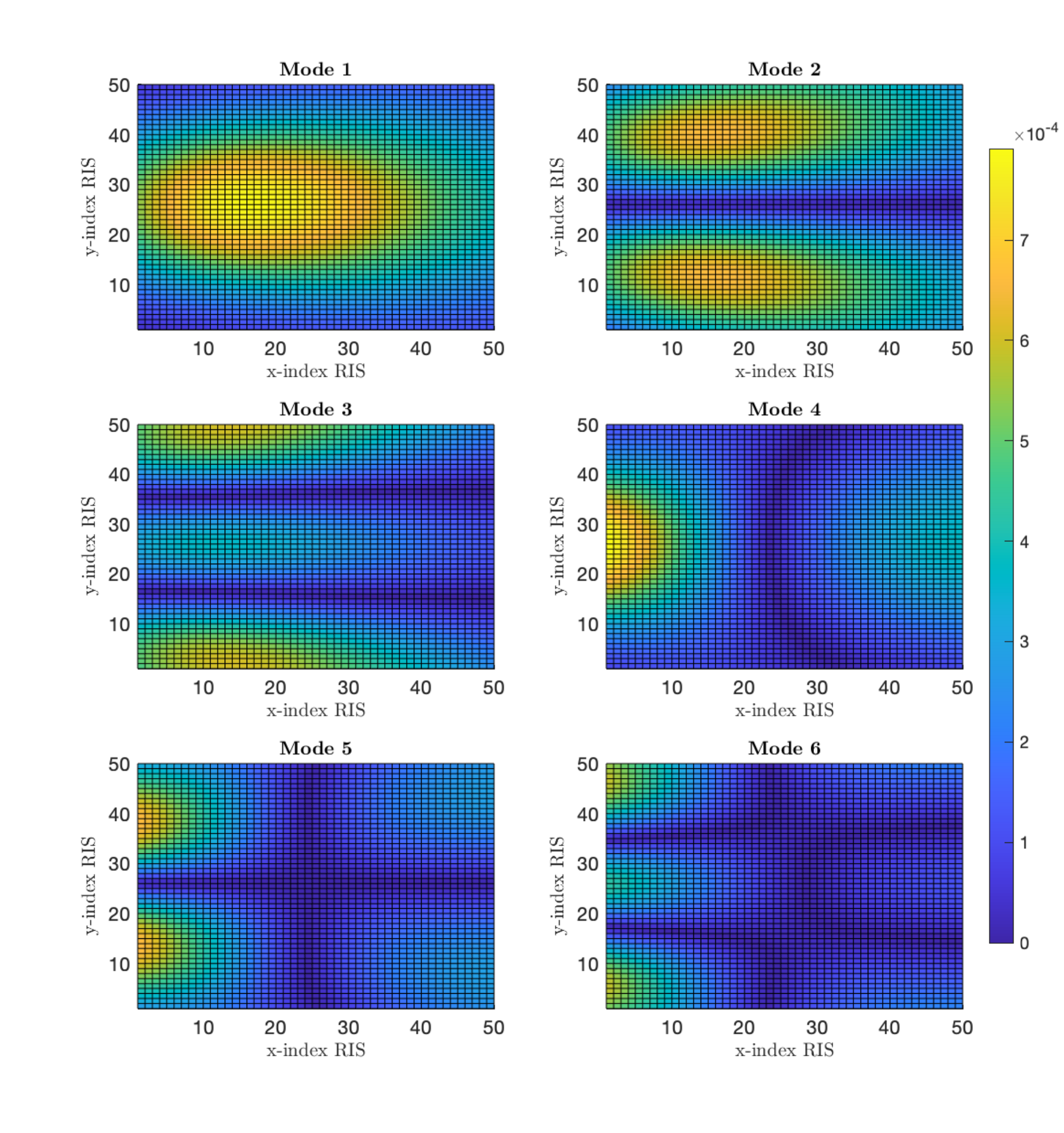}
\caption{Absolute value of the communication modes.}
\end{subfigure}\hfill
\begin{subfigure}{\columnwidth}
\includegraphics[width=0.90\columnwidth]{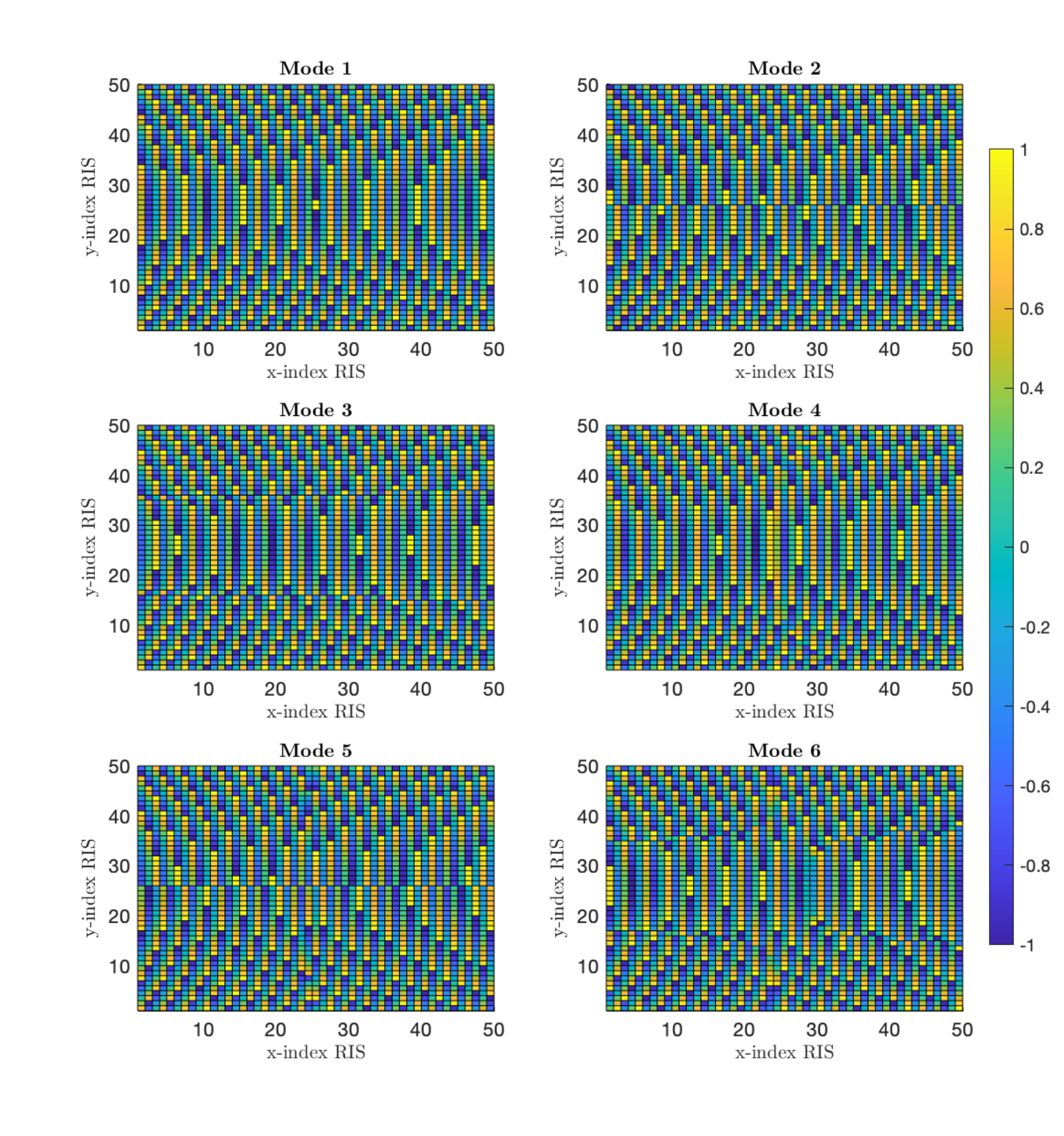}
\caption{Phase (normalized by $\pi$) of the communication modes.}
\end{subfigure}
\caption{Communication modes observed at the RIS.}
\label{fig:modes} \vspace{-0.25cm}
\end{figure*}

In Fig. \ref{fig:modes}, we show $\mathbf{w}_i$ for the six strongest communication modes when the direct channel is blocked and $D=6$ m. 

\begin{figure}[t]
\centering
 \includegraphics[width=0.90\columnwidth]{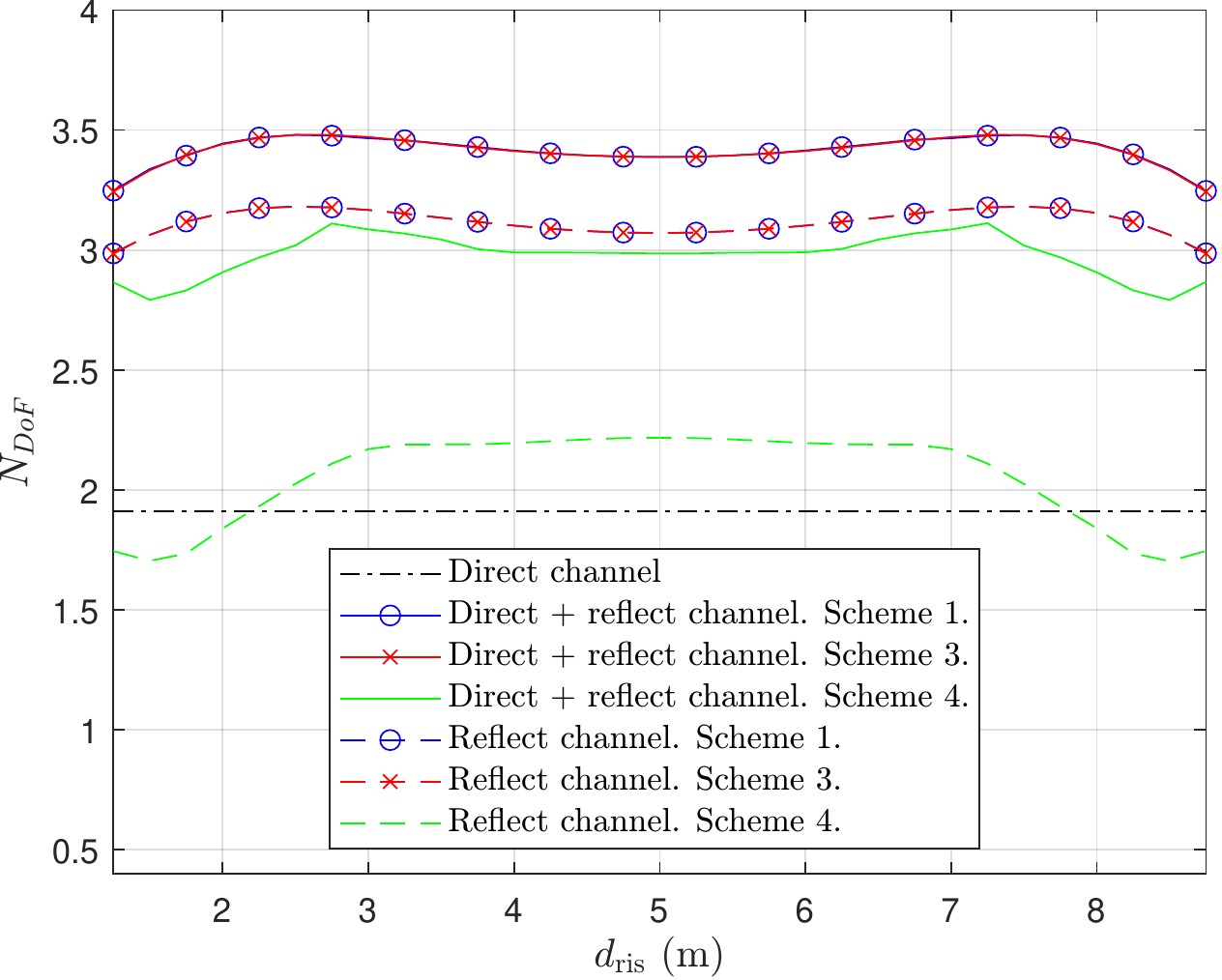}
\caption{Number of \ac{DoF} as a function of the position of the RIS.}
\label{fig:DoF_varydris} \vspace{-0.5cm}
\end{figure}

Finally, we analyze how the position of the RIS impacts the number of \ac{DoF}. This is shown in Fig. \ref{fig:DoF_varydris} for $D=10$ m. We note that the number of effective (strongly coupled) \ac{DoF} fluctuates slowly with the position of the RIS.

\section{Conclusions}
\label{sec:conclusions}
In this paper, we have analyzed the performance of RIS-aided HoloS communications in terms of rate and DoF. We have compared different case studies as a function of the CSI prior knowledge for optimizing the phase shifts of the RIS. The main finding of this paper is that optimizing an RIS based only on location information offers good performance, provided that the channels are predominantly in LoS conditions. This is often the case in high frequency channels, e.g., sub-terahertz and terahertz channels, due to the lack of diffraction.

\bibliographystyle{IEEEtran}
\bibliography{sample}

\end{document}